\documentclass[aps,pre,showpacs,twocolumn,groupedaddress]{revtex4}

\usepackage{graphicx}% Include figure files
\usepackage{dcolumn}% Align table columns on decimal point

\begin{document}

\title{Information weights of nucleotides 
in DNA sequences
%The evidence of Shannon's channel capacity theorem in DNA sequences
}

\author{M.R. Dudek$^1$, S. Cebrat$^2$, M. Kowalczuk$^2$, P. Mackiewicz$^2$, 
A. Nowicka$^2$,\\ 
D. Mackiewicz$^2$, M. Dudkiewicz$^2$}
\affiliation{
$^1$ Institute of Physics, University of Zielona G{\'o}ra,
ul. A. Szafrana 4a, 
PL-65516 Zielona G{\'o}ra, Poland\\
$^2$Department of Genetics, Institute of Microbiology,
University of Wroclaw,
ul. Przybyszewskiego 63/77, PL-54148 Wroclaw, Poland\\
}

\begin{abstract}
The coding sequence in DNA molecule is considered 
as a message necessary to be transferred
to receiver, the proteins, through a noisy information channel
and each nucleotide is assigned a respective information weight.
With the help of the nucleotide substitution matrix we estimated
the lower bound of the amount of information 
carried out by nucleotides which 
is not subject of mutations. 
We used the calculated weights to  
reconstruct $k$-oligomers of genes from the {\it Borrelia burgdorferi}
genome. We showed, that to this aim  
there is sufficient a simple rule, that the number
of bits of the carried information cannot exceed some threshold value.
The method introduced by us is general and applies to every genome.
\end{abstract}


\maketitle

\section{Introduction}
Since the famous paper by Crick et al.\cite{l_Crick} it is 
generally accepted that there exists, in all living organisms, 
a code that makes possible information transfer from the sequences of four
nucleotides in DNA to the sequences of 20 amino acids in proteins.
Namely, the protein sequence is coded by codons  which are
the triplets of nucleotides - each corresponding to one
amino acid in a protein
sequence. There are 64 possible triplets and only 20 amino acids.
Hence, the genetic code is degenerated. Crick et al.\cite{l_Crick} proved, 
that codons are always read from the start, codon after codon, they do not
overlap, there are no
commas between them. Therefore,
each strand of the
coding region of a DNA molecule
could be read in three different reading frames and  
all the statistical analyses of the coding properties
of the DNA sequences should be done in a specific reference system
consistent with the triplet nature of
 the genetic code \cite{l_Dujon0,l_cebrat3,l_Dujon3}.
Any other analysis  basying only on
a single sequence of nucleotides  averages
the coding information of the sequence and the  coding
structure cannot be observed clearly. 
Crick \cite{l_Crick2}, formulated the {\it Central Dogma} 
of molecular biology, stating that genetic information  
first has to be transferred from DNA to RNA and next from RNA to protein.
The wide discussion of the Central Dogma problem can be found  
 in papers by Yockey \cite{l_Yockey1}-\cite{l_Yockey2}.
The genetic code is universal in the sense that 
it is used by all living organisms and there is no found 
a counterexample until now. Thus, there should be the same
simple rule in using the genetic code, 
does not matter how complex the organism is.
 
The coding sequence in DNA
molecules can be thought as a message necessary to be transferred
from source to receiver through a noisy information channel, e.g.
\cite{l_Oliver}, \cite{l_Yockey2}.
Hence, the four letter alphabet (A,T,G,C)
in DNA should  be translated into a   
20 letter alphabet of amino acids in protein.  
Shannon \cite{l_Shannon} considered the generation of 
a message to be a Markov process,
subject to a noise, and he introduced an expression 
for the measure of information  
in a message 
\begin{equation}
H=-k \sum_{i} p_i \log p_i,
\label{entropy}
\end{equation}
\noindent
known as information entropy, 
where $i$ denotes letters of the alphabet under consideration, 
$p_i$ represents the probability of the symbol $i$, and the coefficient
$k$ is for the purpose of a unit of measure. In the case of 
binary alphabet, one usually chooses logarithm base 2 in the above expression.
Then, the amount of information in a message is measured with the help of an
average number of bits necessary to code for 
all possible messages in optimal
way. For example, the field, SEX, in a database needs only one bit to code for
male and female, e.g. $0=$ female and $1=$ male. The question rises, how 
many bits are necessary to measure the amount of information 
carried by  A,T,G,C in DNA coding sequences?
If there is no additional information about the frequency 
of the occurrence of
the nucleotides, then two bits are necessary for each nucleotide. On the other
hand, the codons, coding for amino acids, 
need  six bits of information for each codon, whereas the amino acids
need on average only four bits.
The additional information about the nucleotides, e.g., 
the compositional bias of DNA leading
strand and lagging strand, makes possible to optimize the code in the way that
the most frequent nucleotides can be represented by the 
smaller number of bits than
the less frequent ones. However, there is an essential problem of the stability
of such code with respect to noise (e.g., during translation process).  
This is the case if the one-bit code representing the field SEX
in databases had been accidentally substituted by its counterpart
in the process of reading this database. Thus, the amount of information
representing the code under consideration cannot be too small and it 
should depend on the level of noise.
The sense of the Shannon's channel capacity theorem is 
 that if one wants to send a message
from source to receiver, with as few errors as possible, then 
the code of the message should posses some redundancy. 
 The problem is how to find the suitable units measuring the 
genetic code? Yockey \cite{l_Yockey2},
showed that codon cannot transfer 1.7915 bits 
of its six-bit alphabet
to the protein sequence even if there are no errors (noise). 
This is almost $30$\% of the total information carried out by codons.
We showed, that in the case of the {\it B.burgorferi} genome
the lower bound of the 
fraction of nucleotides, which is not subject to mutations 
could represent even about $36$\% of the nucleotide occurrence 
in the genes. Thus, the number of bits, which cannot be transferred, 
could be much larger.  
   
The necessity of the suitable weight given to nucleotides, 
in such a way that 
they be consistent with the amino acid structure,  is notified  also in
papers dealing with DNA symmetry consideration 
in terms of group theory \cite{l_Duplij}.
In the following,
we suggested an example of the information weights given to
nucleotides, 
which is sufficient for statistical
analyses in the scale of whole genome. In particular, we showed that
only one requirement 
on the capacity of the information channel, 
that the number of bits transferred through
the channel in a message ($3k$-oligomer) cannot exceed a suitable threshold value, 
is sufficient to 
reproduce the nucleotide composition in each codon of the
$3k$-oligomers.

The problem discussed by us is closely related 
with designing DNA codes, which is important for 
biotechnology applications, e.g. in storing and retrieving information in
synthetic DNA strands or 
as molecular bar codes in chemical libraries. This has been
discussed recently by Marathe et al.\cite{l_dnacode} (see also very rich 
literature within).

\section{Information weights of nucleotides}

In papers \cite{l_linear1},\cite{l_linear2},\cite{l_linear3}, 
we concluded that 
in natural genome the frequency of occurrence $f_j$ of 
the nucleotides ($j$=A, T, G, C), 
in the third position in codons, 
is linearly related to the respective
mean survival time $\tau_{j}$,
\begin{equation}
f_{j} = m_0~ \tau_{j} + c_0 ,
\label{r_lin}
\end{equation}
\noindent
with the same coefficients, $m_0$ and $c_0$, for each nucleotide.  
The coefficient $m_0$ is proportional to mutation rate $u$, 
experienced by 
the genome under consideration. This means, that in natural 
genome, with balanced  
mutation pressure and selectional pressure, the nucleotide occurrences
are highly correlated. This observation does not contradict to 
the Kimura's neutral theory \cite{l_neutral} of evolution, which assumes the
constancy of the evolution rate, where the mutations are random events,
much the same as the random decay events of the radioactive decay.
Actually, the mutations are random but they are 
correlated with the DNA composition. Thus, the frequency $f_j$ contains
information specific for genome and therefore 
it  seems to be a natural candidate to model
information weight for nucleotides. In this case, 
the entropy (Eq.~\ref{entropy}) of a message 
consisting of symbols A, T, G, C reads as 
\begin{equation}
H=\sum_{j=A,T,G,C} f_j \log_2(\frac{1}{f_j})=<{\log_2(\frac{1}{f_j})}>,
\label{entropy2}
\end{equation}
\noindent
where $\log_2(\frac{1}{f_j})$ represents the number of bits necessary 
to code nucleotide $j$ in optimal way, and 
the brackets in the last expression denote an expectation value
of the number.
 
\begin{figure}[ht]
\begin{picture}(230,150)(180,0.0)
\includegraphics[height=7.5cm,width=6.5cm]{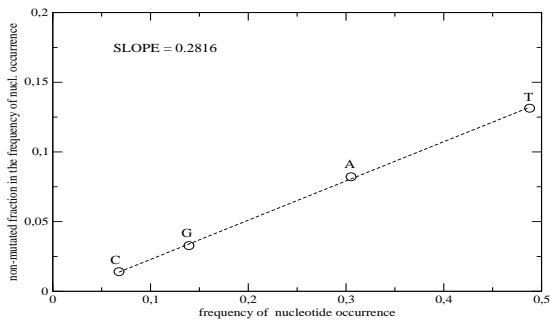}
\end{picture}
\caption{Relation between fraction of non-mutated nucleotides
and their frequency of occurrence in the case of the coding sequences of the
{\it B.burgdorferi} genome. The lower bound for the fraction 
of non-mutated nucleotides is presented when the mutation rate $u=1$. 
}
\label{fig1}
\end{figure}

\begin{figure}[h]
\begin{picture}(100,190)(5,0.0)
\includegraphics[height=8cm,width=7cm]{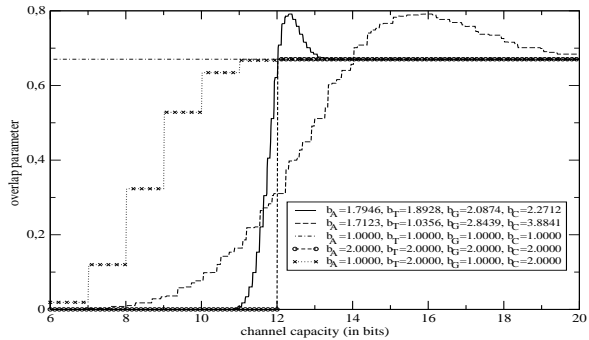}
\end{picture}
\caption{The effect of the choice of the average number 
of bits $B_j$,  representing nucleotides, on pattern recognition.
There is shown the dependence of the overlap parameter $q$ 
on the channel capacity between
computer generated $6$-oligonucleotides and the $6$-oligonucleotides 
in the position $3$ in codons of the {\it B.burgdorferi} genome.
}
\label{fig2}
\end{figure}

The question rises, whether we could 
estimate a fraction $\tilde{f_j}$ 
in the frequency $f_j$, which is not influenced 
by  nucleotide substitutions. To answer the question, we considered 
the probability, that nucleotide $j$ becomes non-mutated, which  is equal to 
\begin{equation}
P_j=1-u\sum_{i \neq j} W_i, 
\label{r_prob}
\end{equation}
\noindent
where $W_i$ is the relative mutation probability, 
\begin{equation}
W_{j}=\sum_{i \neq j} W_{ij},
\end{equation}
\noindent
being a  sum of the relative probabilities 
that nucleotide $j$ will mutate to
the nucleotide $i$, 
and 
$W_A+W_T+W_G+W_C=1$. The parameter $u$ represents mutation rate. 
In the case of the {\it B.burgdorferi} genome, we have 
found (\cite{l_linear1},\cite{l_linear2},\cite{l_linear3}) an empirical
mutation table, applying for genes of leading DNA strand, 
where the  values of $W_{ij}$
 are the following:
 \begin{equation}
\begin{array}{l}
W_{GA}=0.0667~~
W_{GT}=0.0347~~
W_{GC}=0.0470\\
W_{AG}=0.1637~~
W_{AT}=0.0655~~
W_{AC}=0.0702\\
W_{TG}=0.1157~~
W_{TA}=0.1027~~
W_{TC}=0.2613\\
W_{CG}=0.0147~~
W_{CA}=0.0228~~
W_{CT}=0.0350\\
\end{array}
\label{r_wspol}
\end{equation}
\noindent
The probability $W_j$ is related to the mean survival time 
in Eq.~\ref{r_lin} as follows   
(derivation can be found in \cite{l_linear1}): 
\begin{equation}
  \tau_{j}=-\frac{1}{ln(1-u~W_{j})} \approx \frac{1}{u~W_{j}}.
\label{r_approx}
\end{equation}
\noindent
In the extreme case of $u=1$, we can obtain the lower bound for 
the probability  $P_j$ (Eq.~\ref{r_prob}) that 
nucleotide $j$ becomes unchanged by mutation at some instant of time $t$. 
We used these values to construct information weights for 
nucleotides without the contribution of the 
substitutions. To this aim, we normalized the probabilities, $P_j$, 
 and  
 each nucleotide has been assigned  
a value 
\begin{equation}
B_j=\log_2(1/P_j) ,
\label{r_bitlen}
\end{equation}
\noindent
being an average number of bits necessary to code 
this nucleotide in optimal way. We have got the following 
values for the information capacity of the particular nucleotides,
\begin{equation} 
B_A=1.79457, B_T=1.89278, B_G=2.27122, B_C=2.08743.
\label{rB1}
\end{equation}
\noindent
In Fig.~\ref{fig2} there is presented, what is the fraction $\tilde{f_j}$
in the frequency $f_j$ of non-mutated nucleotides when  $u=1$. This fraction
 has been obtained with the help of the 
normalized probabilities $P_j$:
\begin{equation}
\tilde{f_j}=P_j f_j.
\end{equation}
\noindent
In the other extreme case when $u=0$ (no substitutions) the slope of
the corresponding line in Fig.~\ref{fig1} would be equal to $\pi/4$. 
Hence, the lower bound on the fraction of nucleotides, 
which is not subject to mutations 
represents about $36$\% of the nucleotide occurrence 
in the genes. The observed, in Fig.~\ref{fig1},  linear relation between
the fraction $f_j$ of nucleotide in coding sequences and the fraction
$\tilde{f_j}$ of non-mutated nucleotides 
in the fraction is implicated by the linear law in Eq.~\ref{r_lin}.

The numbers in Eq.~\ref{rB1} can be compared 
to the corresponding values 
originating from nucleotide occurrence $f_j$ in the 
position ($3$) in codons of genes. They 
are the following:
 \begin{equation}
B_A=1.7123,B_T=1.0356, B_G=2.8439, B_C=3.8841.
\label{rB2}
\end{equation}
\noindent
Notice, that both series of information weights share 
the same order of appearance. 

\section{Information weights of amino acids}

Recently\cite{l_submit}, 
we showed that the mean survival time of amino acids 
depends on their frequency of occurrence in proteins 
according to a power law,
\begin{equation}
\tau_j \sim F_j^{\alpha},
\label{r_power}
\end{equation} 
\noindent 
where $j$ denotes 20 amino acids. The exponent has a negative value 
 in the case of both selection pressure and mutation pressure, and  
a positive value in the case of pure mutation pressure. 
There is no selectional data for single genome but there is 
available a table of amino acid substitutions published by 
Jones et al. \cite{l_Jones}, which results from statistical analysis
of 16130 protein sequences from few species. 
The table represents so called PAM1 matrix,
corresponding to 1 percent of substitutions between two compared sequences.
We found \cite{l_submit} that 
 $\alpha \approx -1.3$ for the table. 
 In the case of the {\it B.burgdorferi} genome,
 we succeeded to generate a table of substitutions representing a 
 pure mutational pressure 
 applied onto genes from leading DNA strand, and we have got 
 $\alpha \approx 0.3$ \cite{l_submit}. The exponent concerning pure mutational pressure
 is species specific.

 The power law relation in Eq.~\ref{r_power} confirms  
 that the occurrence of amino acids in protein sequences 
 is highly correlated.  
 Therefore, for each amino acid we calculated the probability
 that it is unchanged, in the same way as in Eq.~\ref{r_prob}.
 Next, we calculated the respective information weights. They take the
 values presented in Table~\ref{table1}.

\setlength{\tabcolsep}{4pt}
\begin{table}
\caption{Average number of bits representing 20 amino acids calculated
with the help of the PAM1 substitution table published by Jones et al. \cite{l_Jones}.}
%\begin{center}
\begin{ruledtabular}
\begin{tabular}{|l|l|}
%\hline\noalign{\smallskip}
amino acid & average number of bits \\
\noalign{\smallskip}
\hline
\noalign{\smallskip}
A&	3.9382	\\
R&	3.3455	\\
N&	4.9298	\\
D&	4.9286	\\
C&	4.9177	\\
Q&	4.9273	\\
E&	4.9307 \\
G&	3.9266	\\
H&	4.9230	\\
I&	4.3469	\\
L&	3.3503	\\
K&	4.9301	\\
M&	5.9207	\\
F&	4.9221	\\
P&	3.9233	\\
S&	3.3570	\\
T&	3.9344	\\
W&	5.9157	\\
Y&	4.9228	\\
V&	3.9362	\\
%\hline
\end{tabular}
\end{ruledtabular}
%\end{center}
\label{table1}
\end{table}

\setlength{\tabcolsep}{4pt}
\begin{table}
\caption{The lower and upper bound for the average number of bits representing 
$6$-tuples of nucleotides in position
($1$),($2$), and ($3$) in codons of $18$-oligomers cut off from 
the genes of the
{\it B.burgdorferi} genome, and the $6$-tuples of the corresponding 
amino acids.}
\label{table2}
\begin{ruledtabular}
%\begin{center}
\begin{tabular}{|l|l|l|l|l|}
%\hline\noalign{\smallskip}
& amino acids & nucl.(1) & nucl.(2) & nucl.(3) \\
\noalign{\smallskip}
\hline
\noalign{\smallskip}
lower bound & 20.0778 & 10.7674& 10.7674&10.7674\\
upper bound & 31.5625 & 13.4435& 13.6273&13.0651\\
%\hline
\end{tabular}
\end{ruledtabular}
%\end{center}
\end{table}

\section{Model of information flow through a channel}

In our case, the counterpart of the Shannon's message in a channel
is a $3k$-oligomer of nucleotides and the sequence is translated 
into corresponding $k$ amino acids.  
We analyzed the case of $k=3, 4, \ldots, 10$.
  The first step, we have done, was the partitioning of all the genes 
under consideration into non-overlapping $3k$-oligomers. 
Next, each nucleotide
has been assigned information weight $B_j$ (Eqs. \ref{r_bitlen} and 
\ref{rB1}) 
according to the description
in the previous section. 

We analyzed all possible 
$3k$-oligomers, for each gene, and we 
found the 
lower bound and the upper bound
of the information capacity 
\begin{equation}
 b_k=\sum_{i=1}^k B_{j_i} 
 \label{r_blok}
\end{equation} 
\noindent
of $k$-nucleotides separately in the position 
($1$), ($2$), and ($3$) in codons. 
It is known that the three subsequences of nucleotides, in the three 
positions in codons, 
 are highly correlated and their composition 
is strongly asymmetric  (e.g., \cite{l_Cebrat2},\cite{l_Cebrat4}).
The triplets of nucleotides (codons) have been
translated into amino acids and the lower and upper bounds have been
found also for amino acids. The results for $k=6$ 
($18$-oligomers have been
considered) are presented in Table~\ref{table2}, where the weights $\{B_j\}$ 
which 
have been used  
are defined as in Eq. \ref{rB1}.
Notice, that the range of the differences among the numbers 
of bits representing 
oligonucleotides specific for each position in codons of the 
examined $18$-oligomers is of the order of maximum $3$ bits whereas the
corresponding range for the amino acids is about four times larger.
We would like to underline that if we had measured the nucleotides 
with the values of $B_j$ introduced in Eq.~\ref{rB2}, the ones 
originating from the nucleotide occurrences
$f_j$, then the range of the corresponding differences 
would be almost four times
larger. This is trivial: the increase in the amount of  the 
carried information implies larger flexibility of information packing
and it increases its stability with respect to substitutions. It is worth
to add that contrary to the information weights for nucleotides 
there is almost no difference in the amount of 
information carried by amino acids between the case of the 
weights derived from
the PAM1 matrix  and the case of weights derived from the frequency 
of occurrence of the particular 
amino acids. 
 The  result suggests that only a fraction of 
 the coding sequence in DNA molecule is considered to be important 
  for the synthesis
 of proteins and the fraction   
could have very small packing flexibility. 
The tight packing of the nucleotide information would be consistent  with
our earlier observation that three subsequences of nucleotides, 
representing nucleotides in position 1, 2, and 3 of codons, 
have strongly asymmetric composition 
(e.g., \cite{l_Cebrat2},\cite{l_Cebrat4}). 

\begin{figure}
\begin{picture}(100,190)(5,0.0)
\includegraphics[height=8cm,width=7cm]{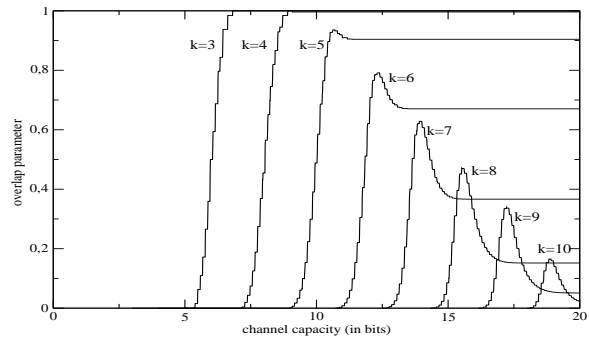}
\end{picture}
\caption{Dependence of the overlap parameter 
on the channel capacity between
computer generated $k$-oligonucleotides and the $k$-oligonucleotides 
in the position $3$ in codons of the {\it B.burgdorferi} genome, when
$k=3, 4, 5, 6, 7, 8, 9, 10$.
}
\label{fig3}
\end{figure}

 One of our main results,
is that in the case of suitably chosen information weights $B_j$
for nucleotides, the only one requirement,
that the number of bits in $3k$-oligomer
cannot exceed some threshold value,
is sufficient to
reproduce the nucleotide composition in each codon of the
$3k$-oligomers. The premise of the property could be seen in Fig.~\ref{fig2},
where an  overlap parameter is plotted between the classes of the
$6$-nucleotides
used in the third position in codons of genes of the {\it B.burgdorferi} and
these $6$-oligonucleotides from among all $4096$ $6$-oligomers
(there are $4^k$ $k$-tuples in the four letter alphabet),
which fulfill the condition that
the total number of bits cannot exceed an assumed
threshold value (information
channel capacity).
 The overlap parameter
has been defined as follows:
\begin{equation}
q=\frac{x}{x+y+z}
\end{equation}
\noindent
where $x$ denotes the number of generated classes of the
sequences, $k$ nucleotides long, which are exactly the same $k$-sequences
as in
natural genome, $y$ denotes the number of generated classes of sequences,
which are different from those in
natural genome, and $z$ denotes  the number of classes of
sequences in natural genome which
have not been selected by the condition that the
total number of transferred bits cannot
exceed some assumed value. In Fig.~\ref{fig2}, we can notice that
the overlap parameter can take maximum value if the average number of bits
representing nucleotides is suitably chosen. There is no recognition
of the compared $k$-oligonucleotides in the case of the loss of information
when all nucleotides have assigned the same weight $B_j=1$.
In the case when all nucleotides have assigned a weight equal to two bits
there is only a signal that below same threshold value of
assumed channel capacity
there is no coincidence between natural and artificial
sequences. There  are, in the figure,  two
 curves with the largest value of maximum value of $q$,
 which are corresponding to two different sets $\{B_j\}$ of information weights
 associated with nucleotides. The left curve represents information
 capacity
calculated with the help of the
empirical substitution table of
nucleotides \cite{l_linear1},\cite{l_linear2},\cite{l_linear3}
(Eq.~\ref{r_wspol}),
whereas
the right one is related with the frequency of occurrence of nucleotides
in genome.
 In general, one could find more representations $\{B_j\}$ of
 information weight
 leading to the same result (the same maximum value). The trivial ones are those
  representations which correspond to other values of mutation rate $u$.
 We expect that the optimum representation
  should be that which
 requires less amount of redundant bits. Hence, the left curve
in Fig.~\ref{fig2} could represent the case with the minimal
information weights for nucleotides necessary to be transferred to proteins.
The results for $k$-oligonucleotides with another value of $k$ can be found in
Fig.~\ref{fig3} in the case of when the information weights
have been chosen as in Eq.~\ref{rB1}.

\begin{figure*}
\begin{picture}(100,190)(100,0.0)
\includegraphics[height=6cm,width=6cm]{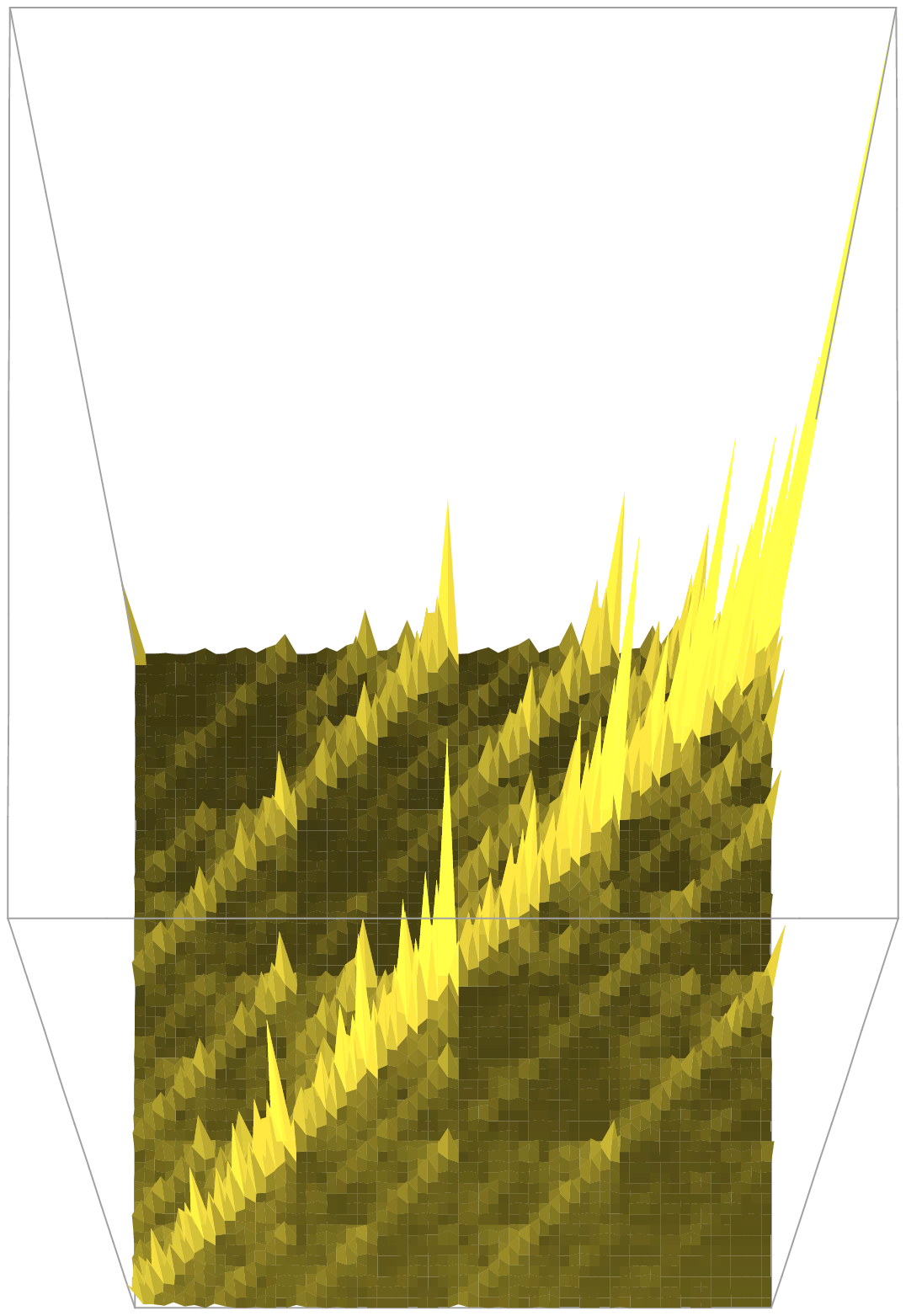}
\includegraphics[height=6cm,width=6cm]{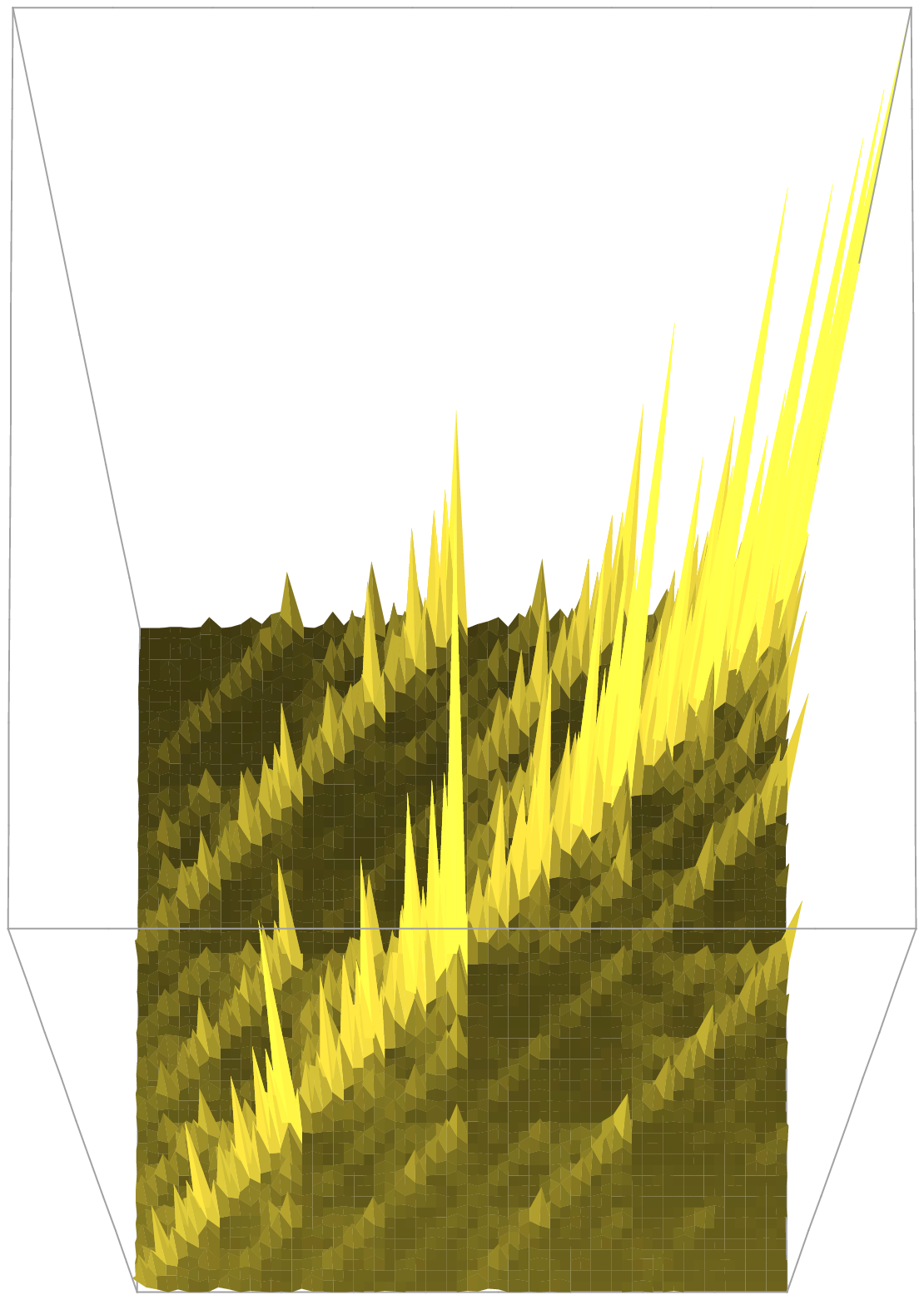}
\end{picture}
\caption{Left: distribution of $6$-oligonucleotides in position $3$ in
codons of genes from the
leading strand of the {\it B.burgdorferi} genome in [A,T,G,C] space. Right:
the same for computer generated sequences.}
\label{fig4}
\end{figure*}

\begin{figure*}
\begin{picture}(100,190)(100,0.0)
\includegraphics[height=6cm,width=6cm]{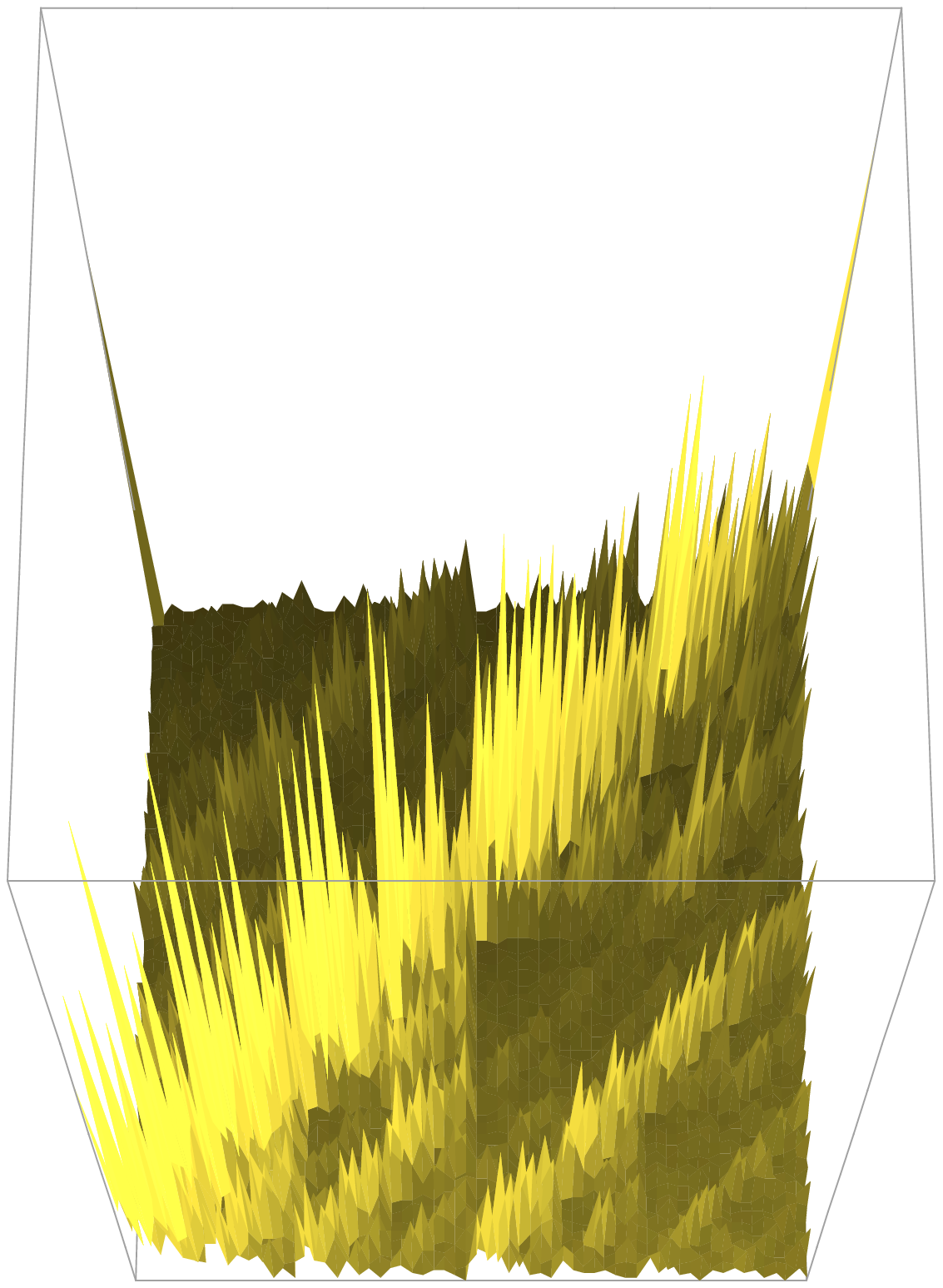}
\includegraphics[height=6cm,width=6cm]{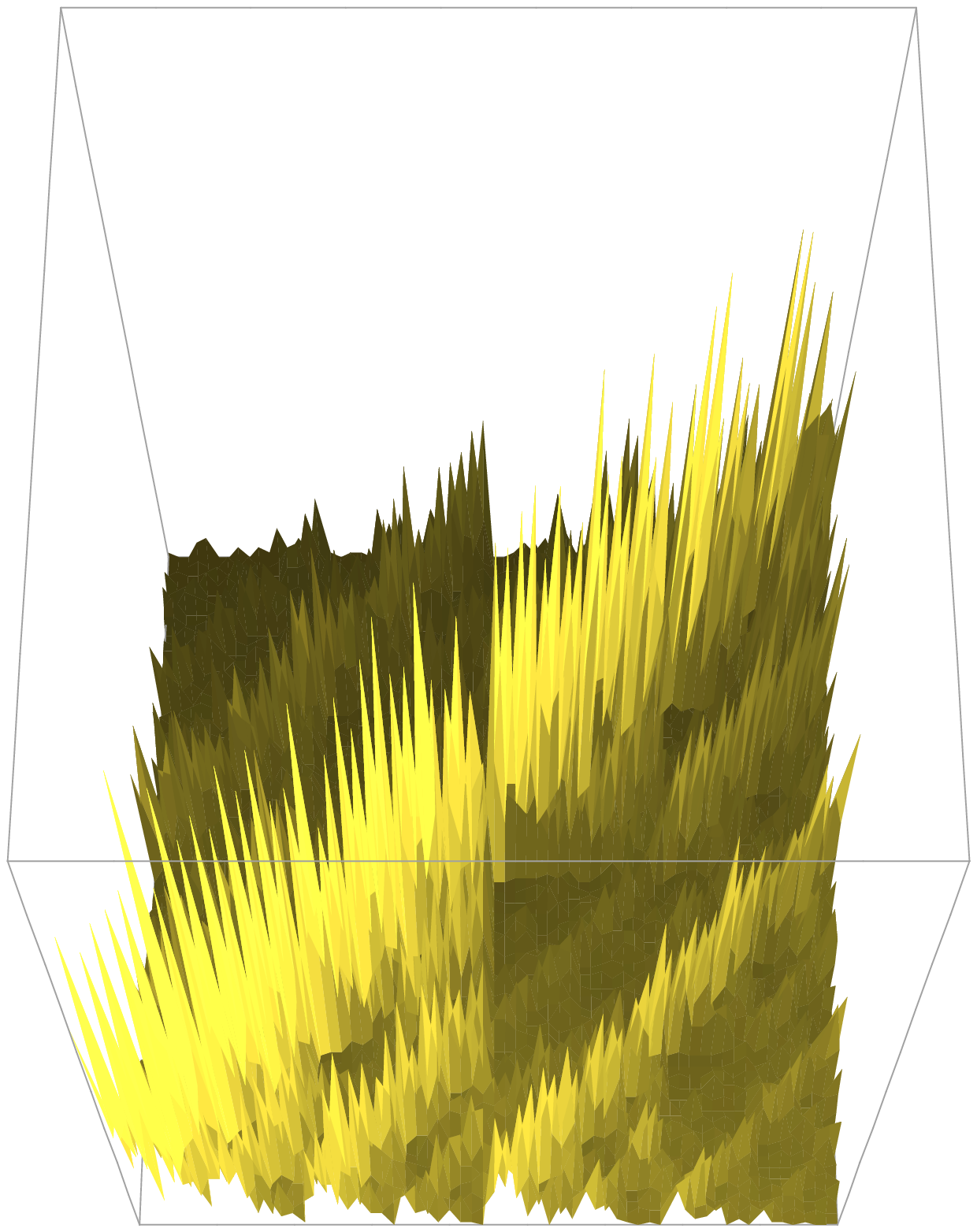}
\end{picture}
\caption{The same as in Fig.~\ref{fig4}, but for position $2$ in codons.
}
\label{fig5}
\end{figure*}

\begin{figure*}
\begin{picture}(100,190)(100,0.0)
\includegraphics[height=6cm,width=6cm]{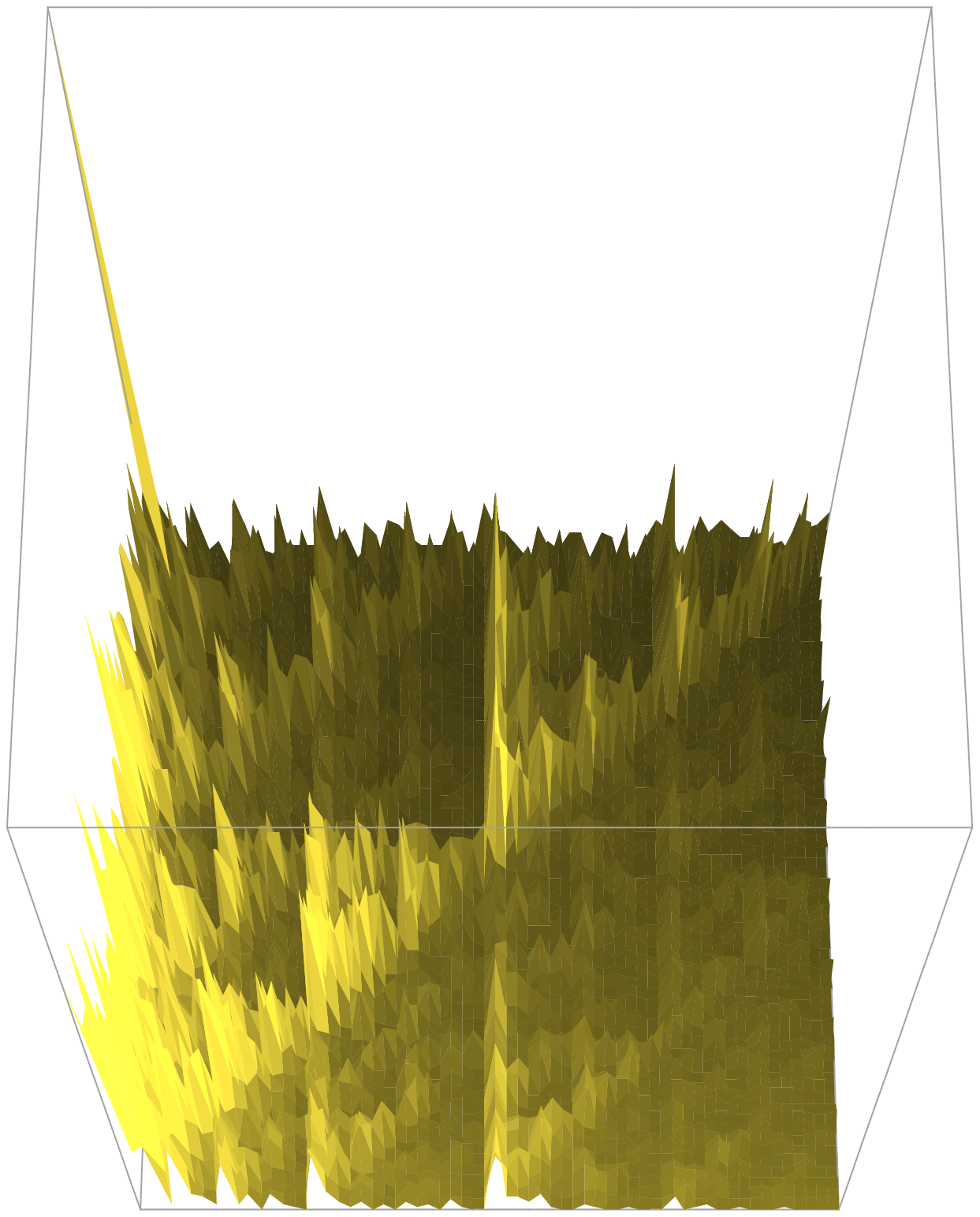}
\includegraphics[height=6cm,width=6cm]{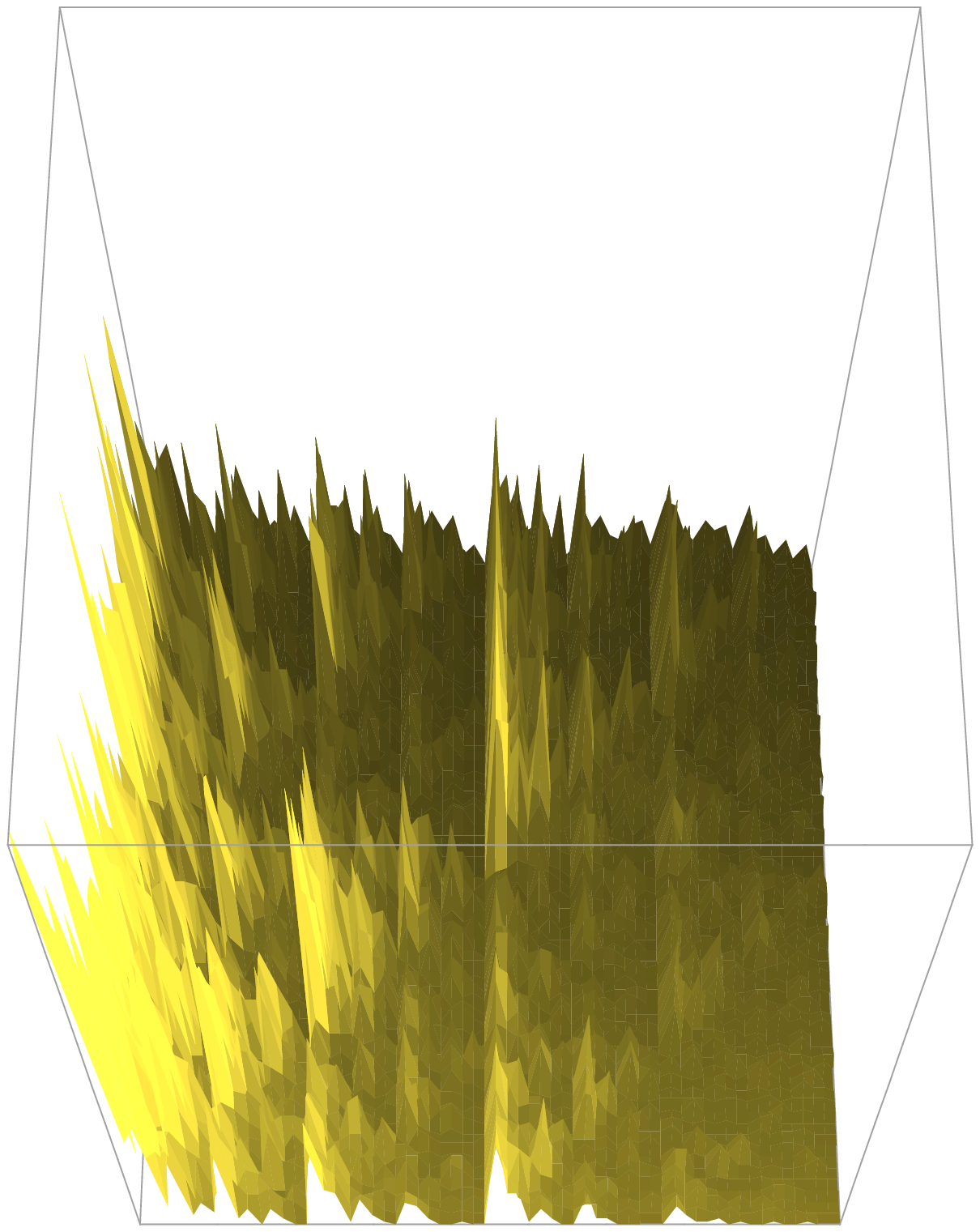}
\end{picture}
\caption{The same as in Fig.~\ref{fig4}, but for position $1$ in codons.
}
\label{fig6}
\end{figure*}

\section{Discussion of the results}
The presence of the lower and upper bound  for information packing both
in DNA sequences and protein sequences imposes natural boundaries
in the model of Shannon's channel. Therefore we generated
with the help of computer
random number generator as many of $3k$-oligomers as there are
 in genes
of natural genome and all the oligomers were fulfilling
the following three conditions:
\begin{itemize}
\item[-] the frequency of occurrence of nucleotides was the same
as in coding sequences of natural genome, separately in position ($1$), ($2$)
and ($3$) in codons,
\item[-] each nucleotide is assigned a value $B_j$ (Eqs. \ref{rB1},\ref{rB2}),
\item[-] the lower bound and the upper bound for the selected
$k$-nucleotides in each position in codons of $3k$-oligomers could not
exceed the values of the lower and upper bound for genes in natural
genome as well
as the triplets of nucleotides from $3k$-oligomers could not exceed the lower
and the upper bound for amino acids after translation of
the considered sequence of nucleotides into a sequence of amino acids.
\end{itemize}

We found, that the distributions of the generated by computer
$k$-nucleotides in all nucleotide positions in codons
 coincide up to the noise introduced by
the over-represented and under-representated oligonucleotides
with the corresponding distributions in natural genome.
This could be seen in the Figs. \ref{fig4}-\ref{fig6}, where we
placed all generated oligomers and natural ones in a space [A,T,G,C]
with the help of IFS ({\it Iterated Function System})
transformation \cite{l_Barnsley}. In the case of $k=6$,
the points of the
space [A,T,G,C] represent all possible $4096$ classes of $k$-tuples
of nucleotides and the hills represent the numbers of the same sequences
in the class.  The detailed construction of the
[A,T,G,C] space can be found in our paper \cite{l_konfer}.
The representation is similar to the chaos game
representation of DNA sequences in the form of fractal images
first developed by Jeffrey
\cite{l_jeffrey} and followed by others, e.g., \cite{l_Hao}.
The size of the hills is closely related to the mutation
pressure and selection. The particular case of the
statistical properties of
short oligonucleotides have been discussed recently  by Buldyrev et al.
\cite{l_buldyrev2}. In particular,
they showed that the number of dimeric tandem repeats in
coding DNA sequences is exponential, whereas in non-coding
sequences it is more often described by a power law.
Other analysis of the $k$-oligomers, like Zipf analysis, can be found
elsewhere, e.g., \cite{l_stan3},\cite{l_ausloos1}.

\begin{figure}
\begin{picture}(100,170)(5,0.0)
\includegraphics[height=8cm,width=7cm]{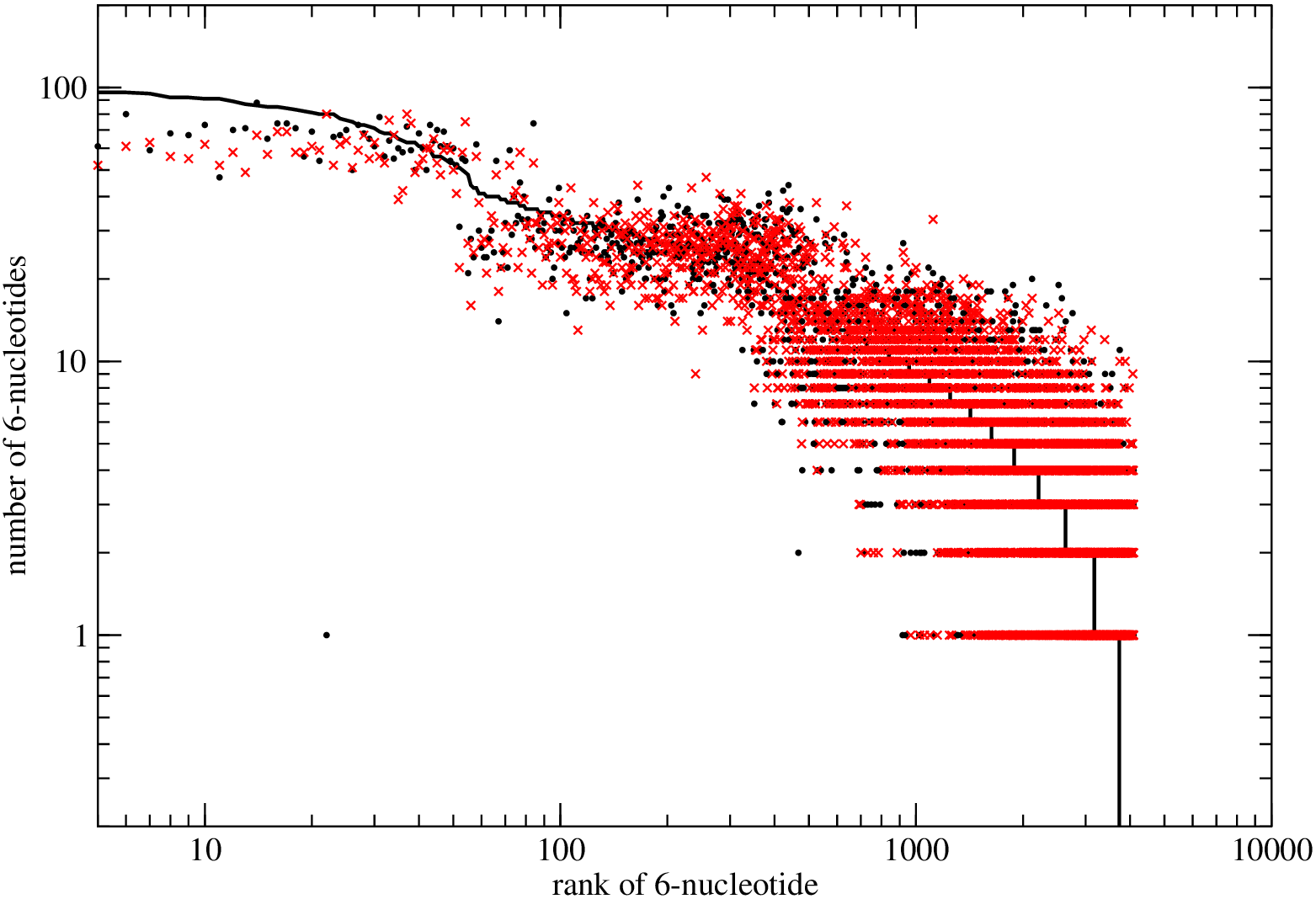}
\end{picture}
\caption{Number of representants in $4096$ classes of $6$-sequences
sorted with respect to their rank number in the case of : $6$-nucleotides
originating from position ($2$) in codons of the {\it B.burgdorferi}
genome (the hills in the left side Fig.~\ref{fig5}),  $6$-nucleotides
generated by computer (the hills in the right side Fig.~\ref{fig5}) where
information weights are defined in Eq.~\ref{rB1}, $6$-nucleotides
generated by computer, where
information weights are defined in Eq.~\ref{rB2}.
}
\label{fig7}
\end{figure}

In the Figs. \ref{fig4}-\ref{fig6},
the best
result with respect to comparison of the natural sequences
with the reconstructed oligomers we have got for
the third position in codons, whereas the worst one we have got
for the second position in codons.
However, the reconstruction of the
second position in codons and the first one also shares many features
common with the natural genome.
The choice of another set of $\{B_j\}$, the one originating
from the nucleotide occurrence
$f_j$ (Eq. \ref{rB2}), leads to very similar results.
We showed this in Fig.~\ref{fig7} for position ($2$) in codons.
In the figure, all $k$-oligonucleotides
have been assigned a rank with respect to their occurrence and
there is plotted their number in each number $4096$ classes 
versus the rank. The three cases are
plotted in the figure: the number of $k$-nucleotides  in natural genome
(continuous line), the number of $k$-nucleotides in generated
sequences, where the information weights have been defined 
in Eq.~\ref{rB1} (dots),
and  where the information weights have been defined in 
Eq.~\ref{rB2} (crosses).
As we can see, even in the case of the weakest reconstruction of DNA oligomers 
for the position ($2$) in codons, the number of representants of each 
of the $4096$ classes well approximates the corresponding number in natural
genome. Much better approximation we have got for the first position in 
codons and the third one. 
It seems, that suggested by us information weights $B_j$, in Eq.~\ref{rB1},  
basying on the mutation table for nucleotides, 
estimate  information  carried out by nucleotides better 
because they correspond to the smaller size of the information channel.
This could be as in the example, given by Yockey \cite{l_Yockey2}, 
of bar code attached to packages items in stores that permits the cashier to
record the price of the item. Namely, the amount of information 
estimated by us with the help of mutation table for nucleotides represents
the sense code whereas the remaining part of it 
represents the redundant bits. Hence, the frequency of occurrence
of nucleotides in natural genome represents two types of information 
being  compromise between selection and mutation pressure.
It is worth to add, that analogous information redundancy 
 in protein sequences is very small.
 
The obtained by us reconstruction of DNA 
sequences with the help of only one rule, that 
the number of transferred bits cannot exceed some threshold value, 
suggests that the width of information channel 
is the basic mechanism of the information packing in DNA coding sequences.  
This is consistent with the statement, 
that if the genetic code is universal, i.e. concerning all living 
organisms,  then 
it must follow the same simple rules of coding.  

There is different research done 
in the field of combinatorial DNA words design
by Marathe et al. \cite{l_dnacode}, in which they discuss some constrains 
imposed on constructed code, like Hamming 
constraint, free energy constraint etc.
Their paper is one of the papers dealing with the study of biotechnological
applications of DNA information. Our result, basying on 
the usage of the table of substitution rates for nucleotides and amino acids,
 could be also used as a new possibility   
 of the designing the  code of synthetic DNA 
 for biotechnology purposes.

\section{Conclusions}
Our results suggest that genetic code 
imposes very tight packing of the nucleotide 
information in  DNA sequences. There is a fraction in the
 frequency of 
occurrence of nucleotide in coding sequences 
which represents the minimum inherited information 
whereas the remaining part concerns redundant information, ensuring that
the DNA code is stable against mutations. 
This observation is not contradictory with 
the Kimura's neutral theory \cite{l_neutral} of evolution. 
The mutations are random events but they are 
correlated with the DNA composition.

There is possible that our  method of the construction 
of the suitable information weights
 of nucleotides  
could be used to discriminate
genes in DNA sequences.   

There is also possibility to use our method in some 
biotechnological applications
dealing with designing  synthetic DNA code.

\end{document}